\documentclass[english,aps,superscriptaddress,twocolumn]{revtex4}

\usepackage{amssymb}
\usepackage{amsmath}
\usepackage{colordvi}
\usepackage[colorlinks]{hyperref}
\usepackage{amsthm}
\usepackage{subeqnarray}
\usepackage{cases}
\usepackage{enumerate}
\usepackage{graphicx}
\usepackage{epsfig}
\usepackage{epstopdf}
\usepackage{subfigure}
\usepackage{color}
\usepackage{url}
\usepackage{verbatim}
\usepackage{mathrsfs}

\usepackage{braket}
\renewcommand{\vec}[1]{\mathbf{#1}}

\begin{document}

\title{Nonclassical Light from Exciton Interactions in a Two-Dimensional Quantum Mirror} 
\author{Valentin Walther}
\email{valentin.walther@cfa.harvard.edu}
\affiliation{ITAMP, Harvard-Smithsonian Center for Astrophysics, Cambridge, Massachusetts 02138, USA}
\affiliation{Department of Physics, Harvard University, Cambridge, Massachusetts 02138, USA}
\affiliation{Department of Physics and Astronomy, Aarhus University, Ny Munkegade 120, 8000 Aarhus C, Denmark}
\author{Lida Zhang}
\affiliation{Department of Physics and Astronomy, Aarhus University, Ny Munkegade 120, 8000 Aarhus C, Denmark}
\author{Susanne F. Yelin}
\affiliation{Department of Physics, Harvard University, Cambridge, Massachusetts 02138, USA}
\author{Thomas Pohl}
\affiliation{Department of Physics and Astronomy, Aarhus University, Ny Munkegade 120, 8000 Aarhus C, Denmark}

\begin{abstract}
Excitons in a semiconductor monolayer form a collective resonance that can reflect resonant light with extraordinarily high efficiency. Here, we investigate the nonlinear optical properties of such atomistically thin mirrors and show that finite-range interactions between excitons can lead to the generation of highly non-classical light. We describe two scenarios, in which optical nonlinearities arise either from direct photon coupling to excitons in excited Rydberg states or from resonant two-photon excitation of Rydberg excitons with finite-range interactions. The latter case yields conditions of electromagnetically induced transparency and thereby provides an efficient mechanism for single-photon switching between high transmission and reflectance of the monolayer, with a tunable dynamical timescale of the emerging photon-photon interactions. Remarkably, it turns out that the resulting high degree of photon correlations remains virtually unaffected by Rydberg-state decoherence, in excess of non-radiative decoherence observed for ground-state excitons in two-dimensional semiconductors. This robustness to imperfections suggests a promising new approach to quantum photonics at the level of individual photons.
\end{abstract}

\maketitle

\section{Introduction}\label{sec:intro}
The ability to couple light and excitons in a semiconducting material is foundational to the tremendous developments in solid-state optics and nanophotonics research \cite{fryett2018, wang2018}. Exploring the regime of quantum photonics, in which synthetic interactions between photons generate quantum states of light, remains an exciting scientific challenge, since the optical nonlinearity that underlies such interactions is weak in most materials. Remarkable advances have been made by coupling single quantum dots to photonic waveguides \cite{lodahl2017} or by reaching strong exciton-photon coupling in semiconductor microcavities \cite{kavokin2017microcavities}. The latter has revealed a rich phenomenology of nonlinear wave phenomena \cite{baas2004,gippius2004,kasprzak2006bose,shelykh2008,amo2009superfluidity,paraiso2010,amo2010,goblot2019}, while first indications of weak photon correlations have been found only recently in experiments \cite{bloch2018,delteil2019}.

A potential approach to address this challenge and enhance optical nonlinearities in semiconductors is to use excited states of excitons \cite{walther2018giant}. Here, the increased polarizability of excitons in excited Rydberg states leads to greatly enhanced interactions \cite{walther2018interactions,shahnazaryan2017,poddubny2019topological} that can even be large enough to inhibit photon coupling to multiple excitons within mesoscopic distances \cite{walther2018interactions}. As successfully demonstrated in experiments with atoms \cite{murray2016, firstenberg2016, sibalic2018}, this Rydberg blockade mechanism indeed leads to sizable nonlinearities that can be sufficiently large to induce interactions and strong correlations between individual photons. Experiments with Rydberg excitons in Cu$_2$O semiconductors \cite{kazimierczuk2014giant,heckoetter2018,steinhauer2020,Lynch2020,konzelmann2019} have found evidence for Rydberg blockade over distances of up to $5~\mu$m \cite{kazimierczuk2014giant,heckoetter2020,walther2020}, and indeed suggest a substantial enhancement of optical nonlinearities at higher exciton excitation levels \cite{walther2020, zielinska2019}. Yet, nonlinear effects have thus far been confined to the domain of classical optics, largely due to the overall weak exciton-photon coupling \cite{kazimierczuk2014giant} in these systems. 

\begin{figure}[t!]
\begin{center}
 \includegraphics[height=.4\textwidth]{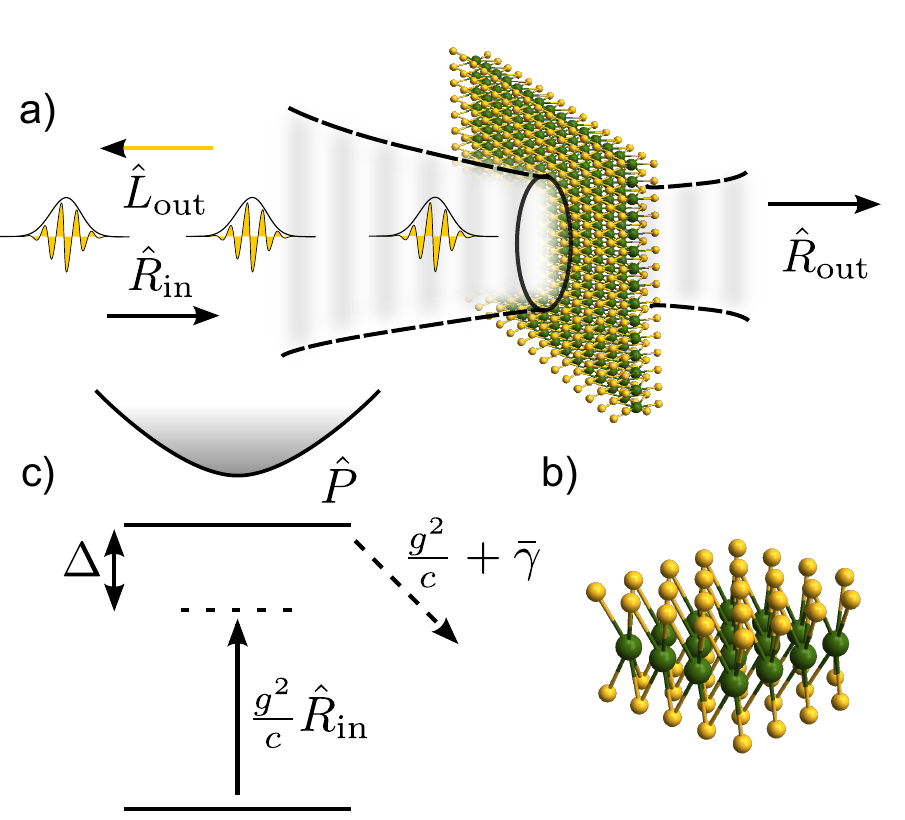}
\end{center}
\caption{a) A quantum light field $\hat{R}_{\rm in}$ impinges on a two-dimensional semiconductor. Under strong driving, its largest fraction is transmitted into $\hat{R}_{\rm out}$, while well-separated photons are back-reflected into $\hat{L}_{\rm out}$. b) Atomistic model of a monolayered transition metal dichalcogenide, where a layer of transition metal atoms (green) is sandwiched between chalcogen atoms (yellow). c) The exciton resonance is described by bosonic operators $\hat{P}(\vec{r})$, coupled to the laser field at a coupling strength $g$ with a detuning $\Delta$. This coupling produces a natural decay rate $\gamma = g^2/c$, which can be elevated by additional decay and dephasing $\bar{\gamma}$.}
\label{fig:fig1}
\end{figure}

Strong light-matter interactions, on the other hand, are possible in a new class of two-dimensional semiconductors, monolayer transition metal dichalcogenides (TMDCs), that has emerged in recent years and offers new perspectives for the manipulation of light, owing to its promising electrooptical properties \cite{lundt2019,glazov2020}. Excitons in this material feature extraordinarily strong coupling to light \cite{wang2018}, and its in-plane translationally invariance renders this coupling highly mode selective. This yields a very effective mirror \cite{zeytinoglu2017,back2018,scuri2018}, where a single layer of the material can reflect light with more that $80\%$ efficiency, limited only by lattice defects and other non-radiative decay mechanisms. While the nonlinearities of TMDC monolayers due to heat diffusion \cite{scuri2018} or collisional interactions between ground-state excitons \cite{katsch2018,barachati2018,ferrier2011,vladimirova2010} are generally small, the possibility to realize a controllable mirror at the smallest possible scales offers exciting perspectives for electro-optics applications \cite{li2020, ross2013electrical, seyler2015} and optomechanics \cite{li2019}.

Here, we explore the combination of finite-range interactions between excitons in excited states and highly coherent light-matter interactions possible in two-dimensional semiconductors. We analyze the resulting photonic nonlinearities in this system by solving the correlated quantum many-body dynamics of laser-driven excitons with strongly interacting excited states. We identify conditions that afford a mapping to an isolated saturable emitter, which provides an intuitive understanding of the generated photon-photon correlations in the transmitted and reflected light. Remarkably, we find a sizeable antibunching of transmitted photons for surprisingly large dissipation of the interacting Rydberg state, allowing decay rates that can approach and even exceed measured linewidths of ground-state excitons in TMDC monolayers. This robust mechanism for effective photon-photon interactions together with the extraordinary optoelectronic properties of TMDC monolayers \cite{wang2012electronics} offers a promising outlook for the exploration of quantum photonics applications at the nanoscale. 

\section{Two-dimensional excitons coupled to light} \label{sec:light_matter}
Transition metal dichalcogenides are a class of materials whose chemical composition MX$_2$ contains a transition metal (M) such as Mo or W, and chalcogen atoms (X) such as S, Se, or Te \cite{liu2013three}. Much like graphene, these materials can be isolated into individual monolayers with a hexagonal structure as illustrated in Fig.~\ref{fig:fig1}. However, unlike graphene, their monolayers can be direct semiconductors \cite{splendiani2010emerging, mak2010} and exhibit sizable bandgaps of $\sim1.5$~eV in the optical domain. Their direct exciton resonance gives rise to extraordinarily strong coupling to light, as mentioned above. These remarkable optical properties and associated opto-electronics applications have motivated broad explorations into coupled spin-valley physics \cite{xiao2012coupled}, nonlinear effects such as second-harmonic generation \cite{seyler2015, janisch2014, kumar2013, ribeiro2015} and the effects of radiating defects \cite{greben2020}. 

Here, we consider a TMDC monolayer that is illuminated by a light beam detuned by a frequency $\Delta$ from one of the exciton resonances (see Fig.~\ref{fig:fig1}). In the absence of defects, the in-plane translational invariance of the material prohibits any momentum transfer within the material such that the coupling to excitons preserves the in-plane momentum component of scattered photons. Focusing on the interaction with a paraxial light beam under orthogonal illumination, it is therefore sufficient to consider two counter-propagating modes \cite{zeytinoglu2017,back2018,scuri2018}. Photons in these modes can be described by bosonic operators $\hat{\mathcal{E}}_\rightarrow(\vec{r},t)$ and $\hat{\mathcal{E}}_\leftarrow(\vec{r},t)$ that denote the slowly-varying electric-field envelope of the electromagnetic field and yield the photon density operators, $\hat{\mathcal{E}}_\rightarrow^\dagger\hat{\mathcal{E}}_\rightarrow$ and $\hat{\mathcal{E}}_\leftarrow^\dagger\hat{\mathcal{E}}_\leftarrow$, in each mode.   
In free space, the fields propagate with the speed of light $c$, as described by the paraxial wave equation \cite{scully1997}. They couple to the two-dimensional excitons in the semiconductor, described by the bosonic operator $\hat{P}(\vec{r}_\perp,t)$. Here, $\vec{r}_\perp$ denotes the two-dimensional coordinate within the plane of the TMDC, and we will from now on denote three-dimensional spatial arguments as $\hat{\mathcal{E}}_\leftarrow(\vec{r}_\perp,z,t)$, choosing the $z$-axis to be orthogonal to the semiconductor plane and parallel to the light propagation axis. Within the rotating wave approximation and neglecting transverse beam diffraction, the Hamiltonian describing the excitons, the light modes and their coupling can then be written as
\begin{equation}
\begin{aligned}
 \hat{\mathcal{H}} = -&\Delta \int d\vec{r}_\perp \hat{P}^\dagger(\vec{r}_\perp) \hat{P}(\vec{r}_\perp) \\
 + g &\int d\vec{r}_\perp \hat{P}^\dagger(\vec{r}_\perp) [\hat{\mathcal{E}}_\rightarrow(\vec{r}_\perp,0) + \hat{\mathcal{E}}_\leftarrow(\vec{r}_\perp,0)] + \text{h.c.} \\
 -ic &\int d\vec{r}\hat{\mathcal{E}}^\dagger_\rightarrow(\vec{r}_\perp,z) \partial_z \hat{\mathcal{E}}_\rightarrow(\vec{r}_\perp,z) \\
 + ic &\int d\vec{r}\hat{\mathcal{E}}^\dagger_\leftarrow(\vec{r}_\perp,z) \partial_z \hat{\mathcal{E}}_\leftarrow(\vec{r}_\perp,z), 
\end{aligned} \label{eq:hamiltonian}
\end{equation}
where $g$ denotes the light-matter coupling strength. This yields the following Heisenberg propagation equations for the photon dynamics
\begin{align}
 \partial_t \hat{\mathcal{E}}_\rightarrow(\vec{r}_\perp,z, t) &= -c \partial_z \hat{\mathcal{E}}_\rightarrow(\vec{r}_\perp,z, t) - ig \hat{P}(\vec{r}_\perp, t)\delta(z) \label{eq:prop_right}, \\
 \partial_t \hat{\mathcal{E}}_\leftarrow(\vec{r}_\perp,z, t) &= c \partial_z \hat{\mathcal{E}}_\leftarrow(\vec{r}_\perp,z, t) - ig \hat{P}(\vec{r}_\perp, t) \delta(z) \label{eq:prop_left},
\end{align}
which can readily be solved. As illustrated in Fig.\ref{fig:fig1}a, one can define the outgoing photon fields $\hat{R}_{\rm out}(\vec{r}_\perp, t) \equiv \hat{\mathcal{E}}_\rightarrow(\vec{r}_\perp,L, t)$ and $\hat{L}_{\rm out}(\vec{r}_\perp,t) \equiv \hat{\mathcal{E}}_\leftarrow(\vec{r}_\perp,-L,t)$ that propagate away from the mirror to its left and right at a distance $L$, respectively. Defining equivalent expressions, $\hat{R}_{\rm in}(\vec{r}_\perp, t) \equiv \hat{\mathcal{E}}_\rightarrow(\vec{r}_\perp, -L,t)$ and $\hat{L}_{\rm in}(\vec{r}_\perp,t) \equiv \hat{\mathcal{E}}_\leftarrow(\vec{r}_\perp,L,t)$, for the input fields, and letting the distance $L\rightarrow 0$, the solution of Eqs.~(\ref{eq:prop_right}) and (\ref{eq:prop_left}) yields a simple set of input-output relations \cite{walls2007quantum, zeytinoglu2017}
\begin{align}
 \hat{R}_{\rm out}(\vec{r}_\perp, t) &= \hat{R}_{\rm in}(\vec{r}_\perp, t) -i \frac{g}{c} \hat{P}(\vec{r}_\perp, t) \label{eq:input_output} \\
 \hat{L}_{\rm out}(\vec{r}_\perp, t) &= -i \frac{g}{c} \hat{P}(\vec{r}_\perp, t) \label{eq:Lout}
\end{align}
as well as the total field at the position of the semiconductor
\begin{align}
 \hat{\mathcal{E}}_\rightarrow(\vec{r}_\perp,0,t) + \hat{\mathcal{E}}_\leftarrow(\vec{r}_\perp,0,t) &= \hat{R}_{\rm in}(\vec{r}_\perp,t) - i\frac{g}{c} \hat{P}(\vec{r}_\perp,t) \label{eq:total_field},
\end{align}
where we assumed that the system is only driven from the left, such that $\hat{L}_{\rm in}$ can be omitted. The Heisenberg equation for the exciton operator governed by the Hamitonian (\ref{eq:hamiltonian}) together with Eq.~(\ref{eq:total_field}) gives 
\begin{align}
  \partial_t \hat{P}(\vec{r}_\perp, t) = -ig\hat{R}_{\rm in}(\vec{r}_\perp, t) +(i\Delta-\gamma) \hat{P}(\vec{r}_\perp, t). \label{eq:P_ideal}
\end{align}
where $\gamma=\frac{g^2}{c}$ is the rate of radiative decay of the exciton into the forward and backward propagating photon modes. Under realistic conditions, defects and non-radiative processes lead to additional dissipation and in particular cause scattering into other modes. Experiments show \cite{back2018,scuri2018} that such additional losses can be well accounted for by an phenomenological decay constant $\bar{\gamma}$, such that the exciton dynamics can be described by
\begin{align}
  \partial_t \hat{P}(\vec{r}_\perp, t) = -ig\hat{R}_{\rm in}(\vec{r}_\perp, t) -\frac{\Gamma}{2} \hat{P}(\vec{r}_\perp, t). \label{eq:P_linear_spatial}
\end{align}
with an effective complex decay rate $\Gamma = \left( 2\gamma + \bar{\gamma} - 2i\Delta \right)$. 

In the following, we will assume a typical experimental situation \cite{delteil2019} in which the reflected and transmitted photons are detected in the same transverse mode as the incident field. Denoting this detection mode by $E(\vec{r}_\perp)$, one can project into this mode and define new operators 
\begin{align}
 \hat{\mathcal{O}} =  \int \hat{O}(\vec{r}_\perp) E^*(\vec{r}_\perp) \rm d \vec{r}_\perp, \label{eq:detection_mode}
\end{align}
that describe the occupation of the transverse detection and input mode $E(\vec{r}_\perp)$, with $\int |E(\vec{r}_\perp)|^2 \rm d \vec{r}_\perp=1$. In this way, $\mathcal{I}_{\rm out} = \langle \hat{\mathcal{L}}^\dagger_{\rm out} \hat{\mathcal{L}}_{\rm out} \rangle$ describes the linear density of outgoing detected photons, while $\langle\mathcal{P}^\dagger\mathcal{P}\rangle$ count the number of excited excitons in the spatial mode $E$.
The equations of motion then take the simple form
\begin{align}
 \hat{\mathcal{R}}_{\rm out}(t) &= \hat{\mathcal{R}}_{\rm in}(t) -i \frac{g}{c} \hat{\mathcal{P}}(t), \label{eq:input_output_projected}  \\
 \hat{\mathcal{L}}_{\rm out}(t) &= -i \frac{g}{c} \hat{\mathcal{P}}(t), \label{eq:Lout_projected} \\ 
 \partial_t \hat{\mathcal{P}}(t) &= -ig\hat{\mathcal{R}}_{\rm in}(t) -\frac{\Gamma}{2} \hat{\mathcal{P}}(t). \label{eq:P_linear}
\end{align}
Solving this simple set of linear equations in Fourier space gives the reflection and transmission spectrum
\begin{align}
R(\omega)&=- \frac{2g^2}{c(\Gamma+2i\omega)}\label{eq:R2l}\\
T(\omega)&=1 - \frac{2g^2}{c(\Gamma+2i\omega)}\label{eq:T2l}
\end{align}
of the TMDC. For resonant cw-driving ($\Delta=\omega= 0$), the reflection coefficient $R(0) = -(1+\frac{\bar{\gamma}}{2\gamma})^{-1} \approx -1 + \frac{\bar{\gamma}}{2\gamma}$ is limited only by non-radiative losses. The strong exciton-photon coupling of TMDC monolayers can render radiative processes dominant ($\gamma>\bar{\gamma}$), which has made it possible to realize reflection coefficients of more than 80\% \cite{back2018,scuri2018}. Under such conditions, residual transmission with $T(0)=\bar{\gamma}/(2\gamma+\bar{\gamma})$ and photon losses $1-|R(0)|^2-|T(0)|^2=4\gamma\bar{\gamma}/(2\gamma+\bar{\gamma})^2$ are greatly suppressed. 

The phase, $\phi(\omega)$ of the complex reflection coefficient $R(\omega)=|R(\omega)|{\rm e}^{i\phi(\omega)}$ contains information about the photon-exciton interaction dynamics. For a long input pulse with a spectral width well below $2\gamma+\bar{\gamma}$, the reflected light 
\begin{equation}
\hat{\mathcal{L}}_{\rm out}(t) \approx R(0) \hat{\mathcal{R}}_{\rm in}(t - \Delta\tau) \label{eq:time_delay}
\end{equation}
has a pulse delay of
\begin{equation}\label{eq:Delta_tau}
 \Delta \tau = -\left. \frac{d\phi}{d\omega}\right|_{\omega=0} = \frac{2(\bar{\gamma}+2\gamma)}{ 4\Delta ^2+\left( \bar{\gamma}+2\gamma\right)^2}.
\end{equation} 
We can interpret this time as the characteristic time the photon interacts with the monolayer and is transferred to an excitonic excitation. For a perfect material ($\bar{\gamma} = 0$) under resonant driving ($\Delta = 0$), the delay time or photon-interaction time is expectedly given by the radiative lifetime $\Delta \tau = 1/\gamma$ of the exciton. 

We can use this time to scale the equations of motion by introducing dimensionless times $\gamma t \rightarrow t $ and lengths $(\gamma/c) r \rightarrow  r$. This yields a simple set of linear equations 
\begin{align}
 \hat{\mathcal{R}}_{\rm out} (t) &= \mathcal{R}_{\rm in}(t) - i \hat{\mathcal{P}}(t) \label{eq:2lv_Rout},\\
 \hat{\mathcal{L}}_{\rm out} (t) &= - i \hat{\mathcal{P}}(t) \label{eq:2lv_Lout},\\
 \partial_t \hat{\mathcal{P}} (t) &= -i \mathcal{R}_{\rm in}(t) - \frac{\Gamma}{2}\hat{\mathcal{P}}(t), \label{eq:eom_P_linear}
\end{align}
that relate the dimensionless output fields $\mathcal{R}_\text{out}\rightarrow \frac{g}{c}\mathcal{R}_\text{out}$ and $\mathcal{L}_\text{out}\rightarrow \frac{g}{c}\mathcal{L}_\text{out}$  to the incident light field $\mathcal{R}_\text{in}\rightarrow \frac{g}{c}\mathcal{R}_\text{in}$ with only two remaining parameters $\bar{\gamma}/\gamma$ and $\Delta/\gamma$ that determine the dimensionless width $\tilde{\Gamma}=\Gamma/\gamma$.

\section{Classical optical nonlinearities due to finite-range exciton interactions} \label{sec:classical}
Nonlinear optical processes can alter the above behavior, whereby the absorption of one photon can affect the optical response of the system to additional photons and break the conditions that otherwise lead to perfect reflection. Possible mechanisms that have been investigated include lattice heating due to photon absorption \cite{scuri2018}, phonon coupling and exciton-exciton interactions \cite{zeytinoglu2017}. The latter usually give rise to relatively weak nonlinearities, due to the typically short range of exciton interactions. This length scale is, however, enhanced for excitons in excited states \cite{shahnazaryan2017, walther2018giant}, as observed for TMDCs in \cite{chernikov2014exciton, stier2018, gu2019}. Very highly excited states of excitons have been observed in bulk Cu$_2$O semiconductors \cite{kazimierczuk2014giant} and found to generate large optical nonlinearities due to an excitation blockade of multiple Rydberg excitons \cite{walther2020}. Such a blockade is caused by the van der Waals interaction $U({\bf r})=C_6/r^6$ between excitons. The van der Waals coefficient $C_6\sim n^{11}$ increases rapidly with the principal quantum number $n$ \cite{walther2018interactions}. The resulting energy shift of exciton-pair states can inhibit the excitation of multiple excitons within a blockade radius $R_{\rm bl}$, at distances $r$ for which $U({\bf r})$ exceeds linewidth of the excitation process. Experimental evidence for the increase of the blockade radius with $n$ and actual values of $R_{\rm bl} \sim 25$~nm at $n=2$ in TMDCs \cite{gu2019} suggests that blockade radii close to $R_{\rm bl} \sim 1 \ \mu$m may be achieved for $n=10$ \cite{wang2020ryd}. 

Interactions are included in the exciton propagation equation~(\ref{eq:P_linear_spatial}) according to
\begin{align}
 \partial_t \hat{P}(\vec{r}_\perp) = &-i\hat{R}_{\rm in}(\vec{r}_\perp) - \frac{\tilde\Gamma}{2}\hat{P}(\vec{r}_\perp) \label{eq:P_nonlinear} \\
 &- i \gamma^{-1}\!\!\int d\vec{r}^\prime_\perp U(|\vec{r}_\perp-\vec{r}_\perp^\prime|) \hat{P}^\dagger(\vec{r}_\perp^\prime)\hat{P}(\vec{r}_\perp^\prime)\hat{P}(\vec{r}_\perp), \nonumber
\end{align}
and lead to a correlated exciton dynamics that typically prevents a simple analytical solution as in the previous section. In order to gain some more intuitive insights into the resulting optical response, we first analyze the classical-optics limit for a coherent input field with a constant amplitude $R_{\rm in}=\langle\hat{R}_{\rm in}\rangle$, where $\langle\hat{R}_{\rm in}\hat{P}\rangle = R_{\rm in}\langle\hat{P}\rangle$, $\langle\hat{R}_{\rm in}\hat{P}^\dagger\hat{P}\rangle = R_{\rm in}\langle\hat{P}^\dagger\hat{P}\rangle$, etc. Under these assumptions, Eq.~(\ref{eq:P_nonlinear}) results in an infinite hierarchy of equations for products of exciton operators, which can be truncated at the lowest nonlinear order \cite{walther2018giant} to obtain an exact description of the third-order susceptibility $\chi^{(3)}$ of the reflected field as
\begin{equation}
\begin{aligned}
 &L_{\rm out}(\vec{r}_\perp) = \ \chi^{(1)} R_{\rm in}(\vec{r}_\perp) \\
 &+ \int \rm d \vec{r}^\prime_\perp \chi^{(3)}(|\vec{r}_\perp-\vec{r}^\prime_\perp|) |R_{\rm in}(\vec{r}_\perp^\prime)|^2  R_{\rm in}(\vec{r}_\perp) \label{eq:nonlin_refl},
\end{aligned}
\end{equation}
where $\chi^{(1)} = R(0)$ corresponds to the linear reflection coefficient discussed in the preceding section and the nonlinear susceptibility is given by
\begin{align}
 \chi^{(3)}(\vec{r}_\perp-\vec{r}^\prime_\perp) = \frac{16}{\tilde{\Gamma} |\tilde{\Gamma}|^2} \frac{i U(|\vec{r}_\perp-\vec{r}^\prime_\perp|)}{\Gamma +iU(|\vec{r}_\perp-\vec{r}^\prime_\perp|)} \label{eq:chi_3}.
\end{align}
The third-order susceptibility $\chi^{(3)}({\bf r})$ acts as an effective interaction potential for two photons and its specific form affords a simple interpretation. At large distances, $r$, for which the exciton interaction potential $U(r)\ll |\Gamma|$ remains well below the linewidth, the nonlinear kernel scales as $\chi^{(3)}({\bf r})\sim U(r)$ and directly reflects the exciton interaction in this perturbative regime. However, in the opposite limit $U(r)>|\Gamma|$ for distances $r<R_{\rm bl}$ within the blockade radius, the susceptibility approaches a constant $\chi^{(3)}\approx\frac{16}{\tilde{\Gamma} |\tilde{\Gamma}|^2}=-8\chi^{(1)}/|\tilde{\Gamma}|^2$ that reduces the overall reflection due to the blocking of multi-photon reflection at distances below $R_{\rm bl}$.
The blockade radius $R_{\rm bl} = \left| C_6/\Gamma \right|^{1/6}$ follows directly from the denominator in Eq.~(\ref{eq:chi_3}). Its magnitude relative to the waist of the input beam determines the extent of nonlinear effects. To be specific, we chose a Gaussian mode profile $E({\bf r}_\perp)=e^{-r_\perp^2/2\sigma^2}/(\sqrt{\pi}\sigma)$ and consider cw-driving with an amplitude $R_{\rm in}(\vec{r}_\perp) = \mathcal{R}_{\rm in}E(\vec{r}_\perp)$. Upon projecting onto this mode and carrying out some of the integrals in Eq.~(\ref{eq:nonlin_refl}), one finds for the reflected field
\begin{align} \label{eq:nonlin_refl_beam}
 \mathcal{L}_{\rm out} 
  &= -\frac{2}{\tilde{\Gamma}} \mathcal{R}_{\rm in} + \frac{16}{\tilde{\Gamma}|\tilde{\Gamma}|^2} |\mathcal{R}_{\rm in}|^2 \mathcal{R}_{\rm in} F \left( \frac{\sigma}{R_{\rm bl}}, \varphi \right) \nonumber,
\end{align}
where we have defined the function 
\begin{equation} \label{eq:def_F}
 F \left( \frac{\sigma}{R_{\rm bl}}, \varphi \right) = \left( \frac{R_{\rm bl}}{\sigma} \right)^2 \int_0^\infty \frac{1}{1 \pm i x^6 e^{i\varphi} } e^{-\left( x \frac{R_{\rm bl}}{\sqrt{2}\sigma} \right)^2} x \ {\rm d} x.
\end{equation}
and $\varphi=\arg(\tilde{\Gamma})$ denotes the phase of the complex linewidth, which vanishes for resonant excitation. 
The different signs correspond to repulsive ($+$) and attractive ($-$) exciton interactions, but for simplicity we focus here on the repulsive case. The function $F$ determines the nonlinear reflection and gives a particularly simple expression 
\begin{equation}
 R = -1 + 2 |\mathcal{R}_{\rm in}|^2 F \left( \frac{\sigma}{R_{\rm bl}}, 0 \right). \label{eq:reflection_classical}
\end{equation}
for an ideal monolayer ($\bar\gamma=0$) under resonant driving.

\begin{figure}[t]
\begin{center}
 \includegraphics[height=.4\textwidth]{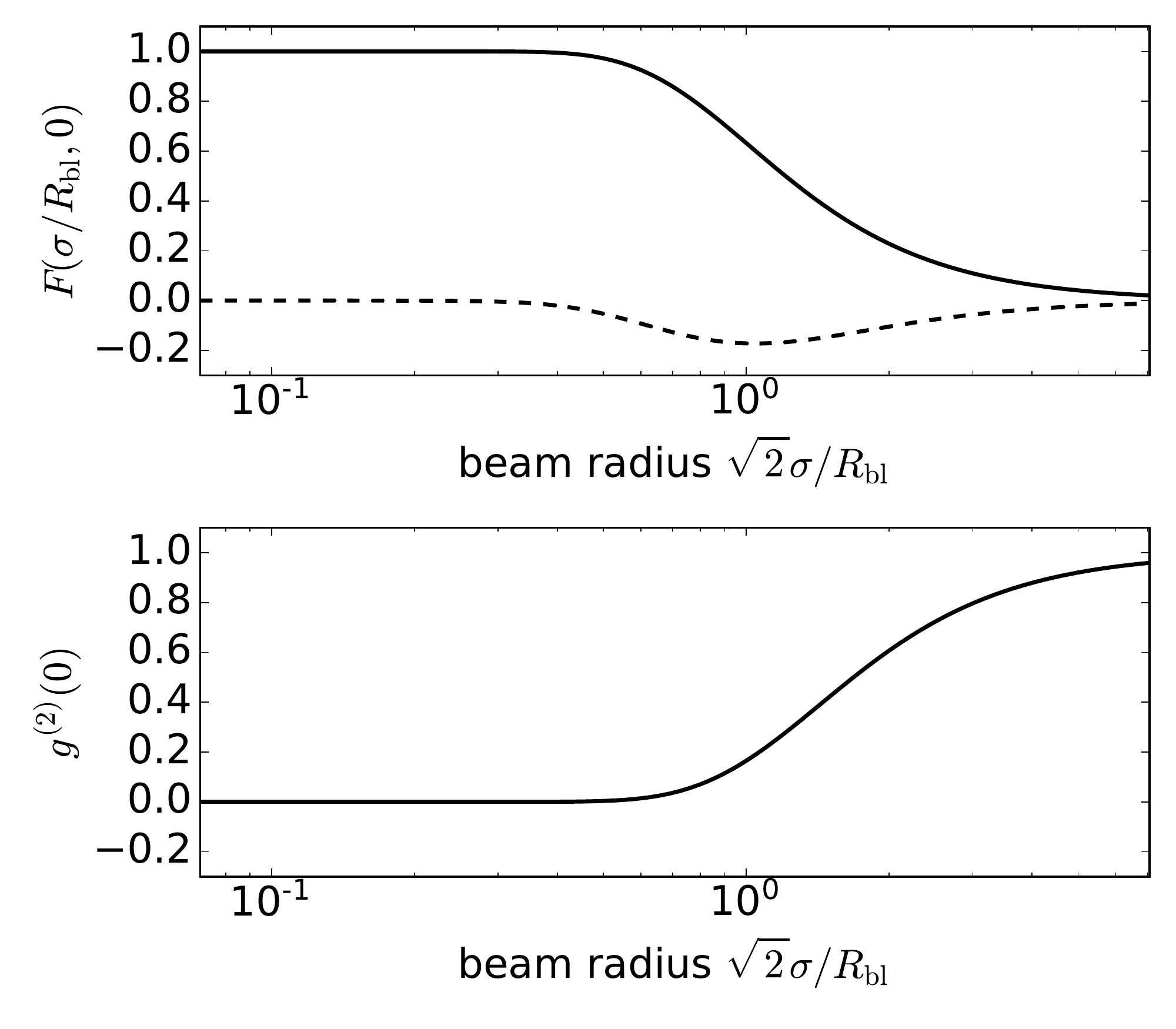}
\end{center}
\caption{Reflection properties for various beam sizes. a) Nonlinear reflection, as characterized by the dimensionless function $F$, reaches a maximal plateau value when $\sigma < R_{\rm bl}$ and falls off for larger beam radii, shown in terms of the function $F$ defined in Eq.~(\ref{eq:def_F}) for resonant driving ($\varphi=0$). The solid line represents the real part (nonlinear reflection), the dashed line represents the imaginary part (nonlinear refraction). b) The equal-time correlation function $g^{(2)}$ of reflected light is strongly suppressed for small beams radii.}
\label{fig:fig2}
\end{figure}
Fig.~\ref{fig:fig2}(a) shows its dependence on the beam radius $\sqrt{2}\sigma$. For large values $\sigma \gg R_{\rm bl}$, multiple excitons can be independently created within the illuminated area $\sim\sigma^2$, which reduces the effects of the nonlinearity and, hence, requires higher light intensities, $|\mathcal{R}_{\rm in}|^2$, to alter the reflection. In this regime, the interactions also affect the shape of the output mode [see Eq.~(\ref{eq:nonlin_refl})], and therefore transfer population out of the incident mode $E({\bf r}_\perp)$. 
This is reflected in the imaginary part of $F$, which increases with decreasing $\sigma$ as the effect of the nonlinearity becomes stronger. Eventually, however, nonlinear losses decrease again once the beam size decreases below the blockade radius $R_{\rm bl}$. In this regime, incident photons only probe the constant inner part of $\chi^{(3)}\approx-8\chi^{(1)}/|\tilde{\Gamma}|^2$, which therefore leads to shape-preserving reflection according to Eq.~(\ref{eq:nonlin_refl}). Concurrently, the nonlinear reflection coefficient saturates to its maximal value, since a single absorbed photon inhibits the reflection for any additional photons in this full-blockade limit $\sigma<R_{\rm bl}$. 
Such strong photon-photon interactions can also give rise to photonic correlations, as we shall discuss in the following.

\section{Quantum states of light}\label{sec:quantm}
Correlations between reflected photons can be quantified by the two-time correlation function 
\begin{align} 
 g^{(2)}_{\rm refl.}(\tau) = \frac{\langle \hat{\mathcal{L}}_{\rm out}^\dagger(t)\hat{\mathcal{L}}_{\rm out}^\dagger(t+\tau)\hat{\mathcal{L}}_{\rm out}(t+\tau)\hat{\mathcal{L}}_{\rm out}(t) \rangle}{\langle \hat{\mathcal{L}}_{\rm out}^\dagger(t)\hat{\mathcal{L}}_{\rm out}(t) \rangle \langle \hat{\mathcal{L}}_{\rm out}^\dagger(t+\tau)\hat{\mathcal{L}}_{\rm out}(t+\tau) \rangle} \label{eq:g2},
\end{align}
which only depends on the time difference between successive photon detection events once the system has reached its long-time steady state ($t\rightarrow\infty$) under cw-driving. Using Eq.~(\ref{eq:2lv_Lout}), $g^{(2)}_{\rm refl.}(\tau)$ can be related directly to temporal correlations of the generated excitons. In particular, we can obtain equal-time correlations ($\tau=0$) from operator products of $\hat{\mathcal{P}}(t)$, using the truncation approach outlined in the previous section. 
As shown in Fig.~\ref{fig:fig2}(b), the obtained dependence of $g^{(2)}_{\rm refl.}(\tau)$ on the widths of the incident laser beam shows similar behavior as discussed above for the nonlinear reflection coefficient. For large values of $\sigma>R_{\rm bl}$, the incident light can simultaneously excite multiple excitons at distances $r>R_{\rm bl}$, which facilitate the simultaneous reflection of multiple photons in the transverse mode $E({\bf r}_\perp)$ such that $g^{(2)}_{\rm refl.}(0)>0$. Eventually, the correlation function approaches $1$ with increasing waist of the incident beam, as multiple excitons can be excited unimpededly for $\sigma\gg R_{\rm bl}$ and, therefore, give rise to uncorrelated photon reflection. However, in the opposite limit of $\sigma<R_{\rm bl}$, the incident light in the driving mode $E({\bf r}_\perp)$ can only generate a single exciton at a time while any further excitation is blocked by the interaction. As a consequence, a single reflected photon effectively blocks reflection of any further light, which passes the monolayer unaffected. The resulting quantum mirror, thus acts as an efficient single-photon filter that generates anti-bunched light with $g^{(2)}_{\rm refl.}(0)=0$, as shown in Fig.~\ref{fig:fig2}(b).

In this strong-blockade limit, in which the interaction $U({\bf r})$ exceeds all other energy scales across the illuminated area, one can simplify Eq.~(\ref{eq:P_nonlinear}) and describe the exciton dynamics by
\begin{align}
\begin{aligned}
 \partial_t \hat{\mathcal{P}} &= -i \mathcal{R}_{\rm in} - \frac{\tilde{\Gamma}}{2}\hat{\mathcal{P}} - i \tilde{U} \hat{\mathcal{P}}^\dagger \hat{\mathcal{P}} \hat{\mathcal{P}}, \label{eq:eom_P}
\end{aligned}
\end{align}
in terms of an effective interaction potential $\tilde{U}={\rm const.}$ that extends over the entire array. Taking the subsequent limit $\tilde{U} \rightarrow \infty$ suppresses all contributions from multiple excitons, and we, for simplicity, consider resonant excitation and neglect additional broadening. This permits truncation of the hierarchy and an adiabatic elimination of the corresponding interaction terms, which leads to a closed set of propagation equations 
\begin{align}
 \partial_t \langle \hat{\mathcal{P}} \rangle &= -i \mathcal{R}_{\rm in} - \frac{\tilde{\Gamma}}{2}\langle \hat{\mathcal{P}} \rangle + 2i \mathcal{R}_{\rm in} \langle \hat{\mathcal{P}}^\dagger \hat{\mathcal{P}} \rangle, \label{eq:2level1} \\
 \partial_t \langle \hat{\mathcal{P}}^\dagger \rangle &= i \mathcal{R}_{\rm in} - \frac{\tilde{\Gamma}^*}{2} \langle \hat{\mathcal{P}}^\dagger \rangle - 2i R_{\rm in} \langle \hat{\mathcal{P}}^\dagger \hat{\mathcal{P}} \rangle, \\
 \partial_t \langle \hat{\mathcal{P}}^\dagger\hat{\mathcal{P}} \rangle &= i \mathcal{R}_{\rm in} (\langle \hat{\mathcal{P}} \rangle - \langle \hat{\mathcal{P}}^\dagger \rangle) - \frac{\tilde{\Gamma}+\tilde{\Gamma}^*}{2} \langle \hat{\mathcal{P}}^\dagger\hat{\mathcal{P}} \rangle\label{eq:2level3}
\end{align}
for the excitons. This simple set of equations describes the dynamics of an effective spin-$1/2$ system with spin operators $\hat{S}_z=\hat{\mathcal{P}}^\dagger\hat{\mathcal{P}}-1/2$ and $\hat{S}_x=(\hat{\mathcal{P}}+\hat{\mathcal{P}}^\dagger)/2$. Similar to so-called Rydberg super-atoms \cite{lukin2001}, in which the Rydberg blockade of an atomic ensemble enables strong photon interactions with a single collective atomic excitation \cite{paris_mandoki2017}, the present setting thus realizes strong coupling between individual photons in a single photonic mode to a single effective saturable exciton. 

\begin{figure}[t!]
\begin{center}
  \includegraphics[height=.5\textwidth]{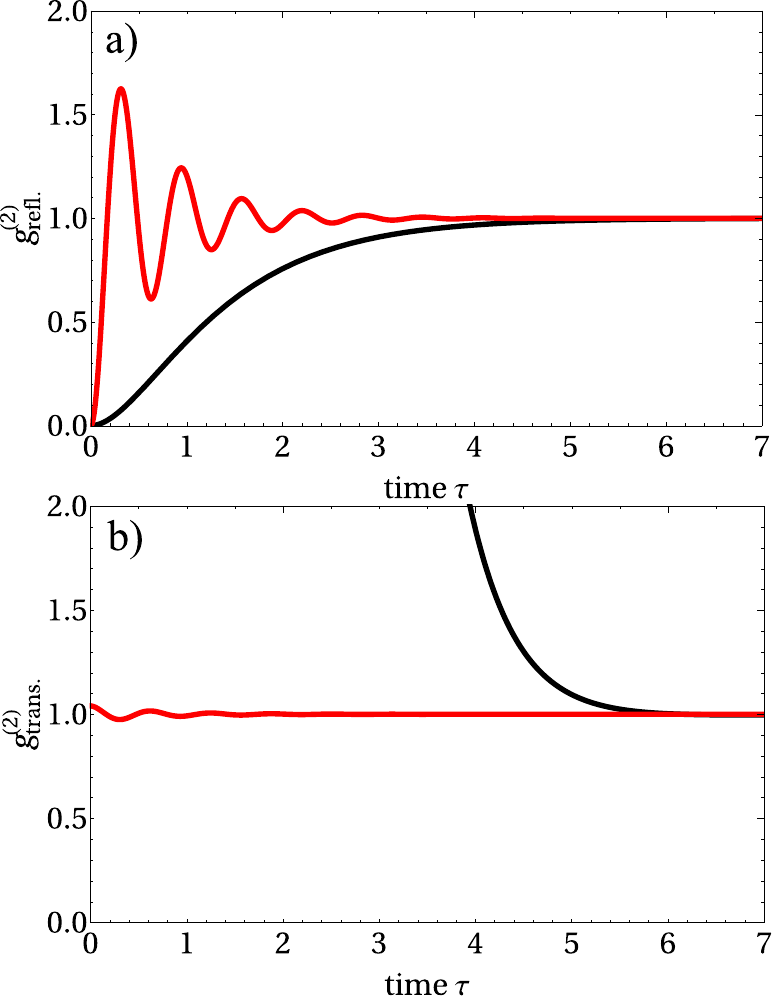}
\end{center}
\caption{Photon correlations in reflection and transmission. a) Photon correlations expressed in $g^{(2)}_{\rm refl.}(\tau)$ for $\mathcal{R}_{\rm in}=0.1$ (black) and $\mathcal{R}_{\rm in}=5$ (red) show strong antibunching and a periodic structure of minima and maxima. b) The transmitted light at the same parameters is strongly bunched under weak driving but exhibits only small oscillations around the long-time limit for strong driving.}
\label{fig:g2_2level}
\end{figure}

On resonance and without additional decay ($\Delta=\bar{\gamma}=0$), the steady-state expectation values
\begin{align}
 \langle \hat{\mathcal{P}} \rangle = -i \frac{\mathcal{R}_{\rm in}}{1+2\mathcal{R}^2_{\rm in}} \qquad \langle \hat{\mathcal{P}}^\dagger \hat{\mathcal{P}} \rangle = \frac{\mathcal{R}^2_{\rm in}}{1+2\mathcal{R}^2_{\rm in}}
\end{align}
yield the nonlinear reflection $\mathcal{L}_{\rm out} = -\mathcal{R}_{\rm in}/(1+2\mathcal{R}^2_{\rm in}) = -\mathcal{R}_{\rm in} + 2 \mathcal{R}_{\rm in}^3+\mathcal{O}(\mathcal{R}_{\rm in}^5)$, in agreement with the third-order result Eq.~(\ref{eq:reflection_classical}) of the previous section. Note that $\langle \hat{\mathcal{P}}^\dagger \hat{\mathcal{P}} \rangle \neq \langle \hat{\mathcal{P}}^\dagger \rangle\langle \hat{\mathcal{P}} \rangle$, which indicates the emergence of photon-photon correlations down to the lowest nonlinear order in the driving field $\mathcal{R}_{\rm in}$. The photon correlation function is readily obtained from Eqs.~(\ref{eq:2level1})-(\ref{eq:2level3}) using the quantum regression theorem, giving the known result 
\begin{align} \label{eq:g2_refl}
 g^{(2)}_{\rm refl.}(\tau) = e^{-\frac{3 \tau}{2}} \left[-\frac{3 \sinh \left(\frac{\tau \sqrt{\kappa}}{2} \right)}{\sqrt{\kappa}}-\cosh \left(\frac{\sqrt{\kappa}}{2} \tau \right) \right]+1
\end{align}
for a driven two-level system \cite{carmichael1976}, where the constant $\kappa = 1-16\mathcal{R}_{\rm in}^2$ is determined by the driving field intensity $\mathcal{R}_{\rm in}^2$. For weak fields ($\kappa > 0$), the pair correlation function monotonically approaches its long-time asymptote $g^{(2)}(\tau) \rightarrow 1$ on a time scale set by the decay rate $\gamma$ (Fig.~\ref{fig:g2_2level}). 
At higher incident intensities for which $\kappa<0$, $g^{(2)}_{\rm refl.}(\tau)$ undergoes damped oscillations with a frequency $\sim\sqrt{-\kappa}$. While the damping time scale is set by the radiative decay rate $\gamma$, the oscillation frequency increases as $\sim R_{\rm in}$, reflecting the coherence of the underlying single-body Rabi oscillations in the limit of strong driving \cite{carmichael1976}. Most importantly, the outgoing light exhibits complete anti-bunching regardless of the driving intensity due to the interaction blockade of simultaneous reflection, as discussed above.

This picture is confirmed by the correlation function 
\begin{align}
 g^{(2)}_{\rm trans.}(\tau) =& -\frac{16 e^{-\frac{3 \tau}{2}} }{(\kappa -1)^2 \sqrt{\kappa}} \left[(\kappa +3) \sinh \left(\frac{\sqrt{\kappa} \tau}{2}\right) \right. \nonumber\\
 &\left. +(\kappa -5) \sqrt{\kappa} \cosh \left(\frac{\sqrt{\kappa} \tau}{2}\right)\right]+1 
\end{align}
 of the transmitted light, described by $\hat{\mathcal{R}}_{\rm out}$. At low intensities, the transmitted light is strongly bunched, with $g^{(2)}_{\rm trans.}(0)$ diverging as $\sim 1/(4\mathcal{R}_{\rm in}^4)$. Since transmission through an otherwise perfectly reflecting monolayer is only possible via exciton-exciton interactions, photons can only be transmitted simultaneously, leading to the large antibunching displayed in Fig.\ref{fig:g2_2level}(b). However, since a single generated exciton blocks reflection of all subsequently incident photons the mirror saturates for high intensities and largely transmits the incident coherent field such that $g^{(2)}_{\rm trans.}(0) \approx 1+1/\mathcal{R}_{\rm in}^2$ quickly approaches unity with increasing driving intensity. 
 
Under strong coherent driving, the nonlinear monolayer, therefore, transmits coherent radiation with weak correlations, while its reflected light is a highly non-classical train of antibunched single-mode photons.
 
\section{Two-photon driving and electromagnetically induced transparency}\label{sec:3level}
\begin{figure}[b!]
\begin{center}
  \includegraphics[height=.2\textwidth]{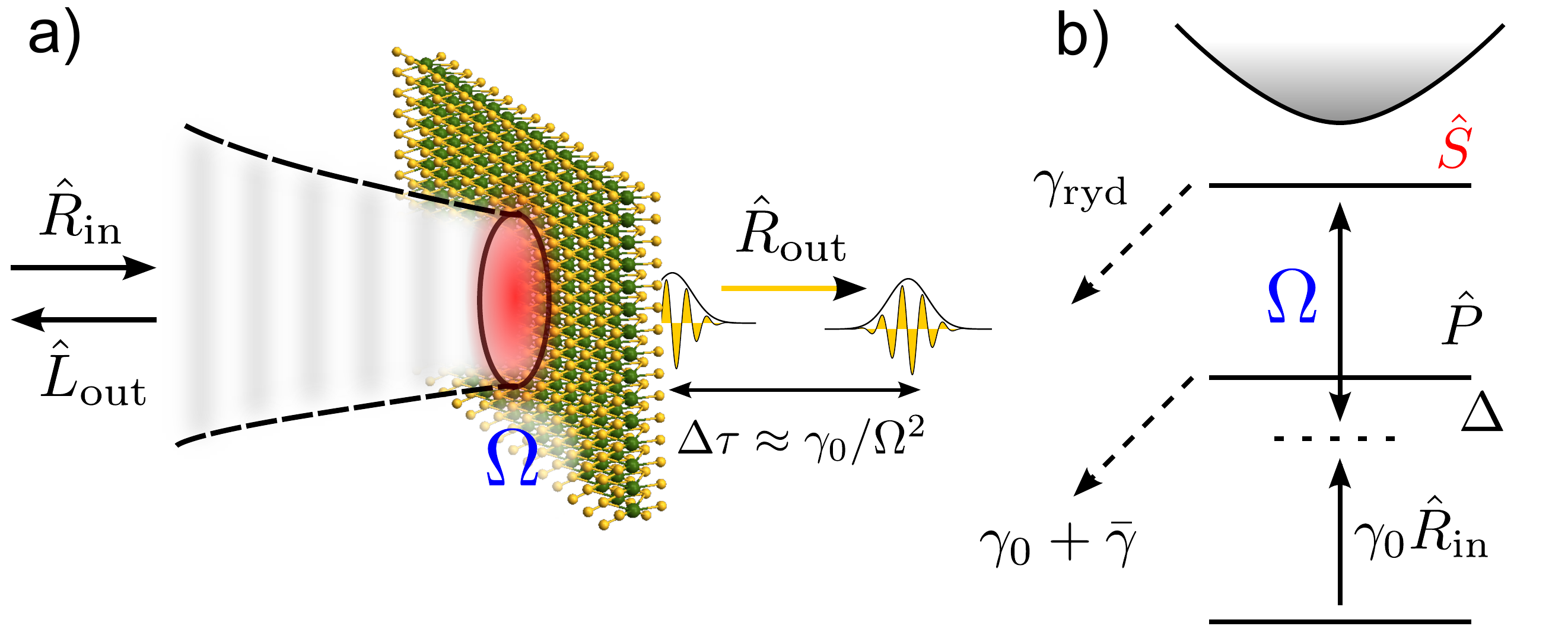}
\end{center}
\caption{Photon dynamics using electromagnetically induced transparency. a) The semiconductor is transparent under conditions of weak driving. Stronger driving breaks EIT and leads to strong reflection off the exciton resonance. The resulting individual photons emerge in reflection with a time separation of $\gamma/\Omega^2$. b) To establish EIT, the exciton resonance is coupled via a second laser to a high-lying Rydberg state, described by $\hat{S}$, at Rabi frequency $\Omega$. The upper state is quasi-stable, limited only by a small total decay rate $\gamma_{\rm ryd}$.}
\label{fig:scheme_3level}
\end{figure}

While the interaction between excitons is enhanced for excited states, their coupling to light ($g$) tends to weaken with increasing principal quantum number $n$. A strong light matter coupling can, however, be maintained by using an additional control laser field. More specifically, this can be achieved via a two-photon coupling of two distinct exciton states as illustrated in Fig.~\ref{fig:scheme_3level}. Hereby, the incident probe field with amplitude $R_{\rm in}$ generates excitons described by the bosonic field $\hat{P}({\bf r}_{\perp})$, as introduced above, while the control laser couples these excitons to a higher lying excited state with Rabi frequency $\Omega$. Denoting the bosonic field operator for these Rydberg excitons by $\hat{S}$, this adds the following Hamiltonian 
\begin{align}
 \hat{\mathcal{H}}_c = & -\delta\int d\vec{r}_\perp \hat{S}^\dagger(\vec{r}_\perp) \hat{S}(\vec{r}_\perp)\\
&\Omega \int d\vec{r}_\perp \left( \hat{P}^\dagger(\vec{r}_\perp) \hat{S}(\vec{r}_\perp) + \hat{S}^\dagger(\vec{r}_\perp) \hat{P}(\vec{r}_\perp)  \right)\nonumber
\end{align}
to the light-matter Hamiltonian introduced in Eq.~(\ref{eq:hamiltonian}), where $\delta$ denotes the total detuning of the two-photon transition to the excited Rydberg state.

Following a similar calculation as in section \ref{sec:light_matter}, we now obtain for the transmission spectrum of the monolayer 
\begin{align}
T(\omega) &= 1+\frac{2 i (\delta -\omega )}{2 \Omega^2+(2 \omega -i \tilde{\Gamma} ) (\delta -\omega)}. \label{eq:Teit} 
\end{align}
While this coincides with Es.~(\ref{eq:T2l}) for $\Omega=0$, a finite control field leads to a vanishing reflection coefficient on two-photon resonance, $\delta=\omega=0$. This is a direct manifestation of electromagnetically induced transparency (EIT) \cite{boller1991, fleischhauer2005}, as has been observed in a range of driven three-level systems \cite{PhysRevLett.84.5094, phillips2003,chen2011}, and can be traced back to the establishment of a dark steady state that does not contain excitons in low-lying states ($\hat{P}$). From Eq.~(\ref{eq:Teit}) we obtain a simple expression 
\begin{align}
  \Delta \tau = \frac{1}{\Omega^2} \label{eq:time_delay_EIT}, 
\end{align}
for the group delay of resonantly transmitted photons, which can now be controlled by the Rabi frequency $\Omega$ and extends well beyond the values given by Eq.~(\ref{eq:Delta_tau}) for the two-level mirror discussed above. This delay $\Delta \tau$ corresponds to the characteristic times for which a transmitted photon is transferred into a Rydberg exciton and, therefore, directly affects the dynamics of the optical nonlinearity. Neglecting the comparably weak interactions between the low lying exciton states, the two-photon resonant dynamics ($\delta=0$) of the excitons  is now described by the coupled equations
\begin{align}
 \partial_t \hat{\mathcal{P}} &= -i \mathcal{R}_{\rm in} - \frac{\tilde{\Gamma}}{2}\hat{\mathcal{P}}  - i \Omega \hat{\mathcal{S}} \label{eq:P} \\
 \partial_t \hat{\mathcal{S}} &= -i \Omega \hat{\mathcal{P}}  -i \tilde{U} \hat{\mathcal{S}}^\dagger \hat{\mathcal{S}} \hat{\mathcal{S}} \label{eq:S},
\end{align}
where we take the limit $\tilde{U}\rightarrow\infty$ to obtain the interaction blockade of multiple Rydberg excitons within the illuminated area of the monolayer. The situation is, however, more complex than in the previous section, since the interaction  does not confine the number of excitons in low lying states. Thus, one has to solve the driven and correlated many-body dynamics of multiple excitons coupled to their strongly interacting excited state. Starting from Eqs.~(\ref{eq:2lv_Rout}), (\ref{eq:P}), and (\ref{eq:S}), this can be expressed in an infinite hierarchy of equations for operator products for the two types of excitons ($\hat{\mathcal{P}}$ and $\hat{\mathcal{S}}$) along with the transmitted photon field ($\hat{\mathcal{R}}_{\rm out}$). Defining the correlators 
\begin{align}
 A_{n,q} &= \langle (\hat{\mathcal{R}}_{\rm out}^\dagger)^n \hat{\mathcal{S}}^\dagger \hat{\mathcal{R}}_{\rm out}^q \rangle, \label{eq:A}\\
 B_{n,q} &= \langle (\hat{\mathcal{R}}_{\rm out}^\dagger)^n \hat{\mathcal{S}} \hat{\mathcal{R}}_{\rm out}^q \rangle, \label{eq:B}\\
 C_{n,q} &= \langle (\hat{\mathcal{R}}_{\rm out}^\dagger)^n \hat{\mathcal{S}}^\dagger \hat{\mathcal{S}} \hat{\mathcal{R}}_{\rm out}^q \rangle, \label{eq:C} \\
 D_{n,q} &= \langle (\hat{\mathcal{R}}_{\rm out}^\dagger)^n \hat{\mathcal{R}}_{\rm out}^q, \label{eq:D}\rangle
\end{align}
this hierarchy can be written in closed form as
\begin{align}\label{eq:3lv_1}
 \partial_t A_{n,q} = &-(n+q) A_{n,q} + \Omega D_{n+1,q} 
 - \Omega \mathcal{R}_{\rm in} D_{n,q} \\
 &+ 2 \Omega \mathcal{R}_{\rm in} C_{n,q}
 - 2  \Omega C_{n+1,q} -  q \Omega C_{n,q-1}  \nonumber
\end{align}
\begin{align}
 \partial_t B_{n,q} = &-(n+q) B_{n,q} + \Omega D_{n,q+1}
 - \Omega \mathcal{R}_{\rm in} D_{n,q}  \\
 &+ 2 \Omega \mathcal{R}_{\rm in} C_{n,q} 
 - 2  \Omega C_{n,q+1} -  n \Omega C_{n-1,q} \nonumber
\end{align}
\begin{align}
 \partial_t C_{n,q} = &-(n+q) C_{n,q} - \Omega \mathcal{R}_{\rm in} ( A_{n,q} + B_{n,q} ) \\
 &+ \Omega B_{n+1,q} + \Omega A_{n,q+1} \nonumber
\end{align}
\begin{align}\label{eq:3lv_4}
 \partial_t D_{n,q} = &-(n+q) D_{n,q} -  n \Omega A_{n-1,q}\\
 &-  q \Omega B_{n,q-1} \nonumber ,
\end{align}
where $D_{0,0} = 1$ and $\Delta=\bar{\gamma}=0$ has been assumed for simplicity. For any finite input power $|\mathcal{R}_{\rm in}|^2$, the solution of this set of equations converges for sufficiently large coefficient matrices $A_{n,q}$, $B_{n,q}$, $C_{n,q}$, and  $D_{n,q}$. We can thus calculate the steady-state expectation values and use the quantum regression theorem to determine the two-photon correlation functions from Eqs.~(\ref{eq:3lv_1})-(\ref{eq:3lv_4}) for a finite set of equations with $n,q<\nu_{\rm max}$, and subsequently verify convergence of the result with respect to $\nu_{\rm max}$.

\begin{figure}[t]
\begin{center}
  \includegraphics[height=.5\textwidth]{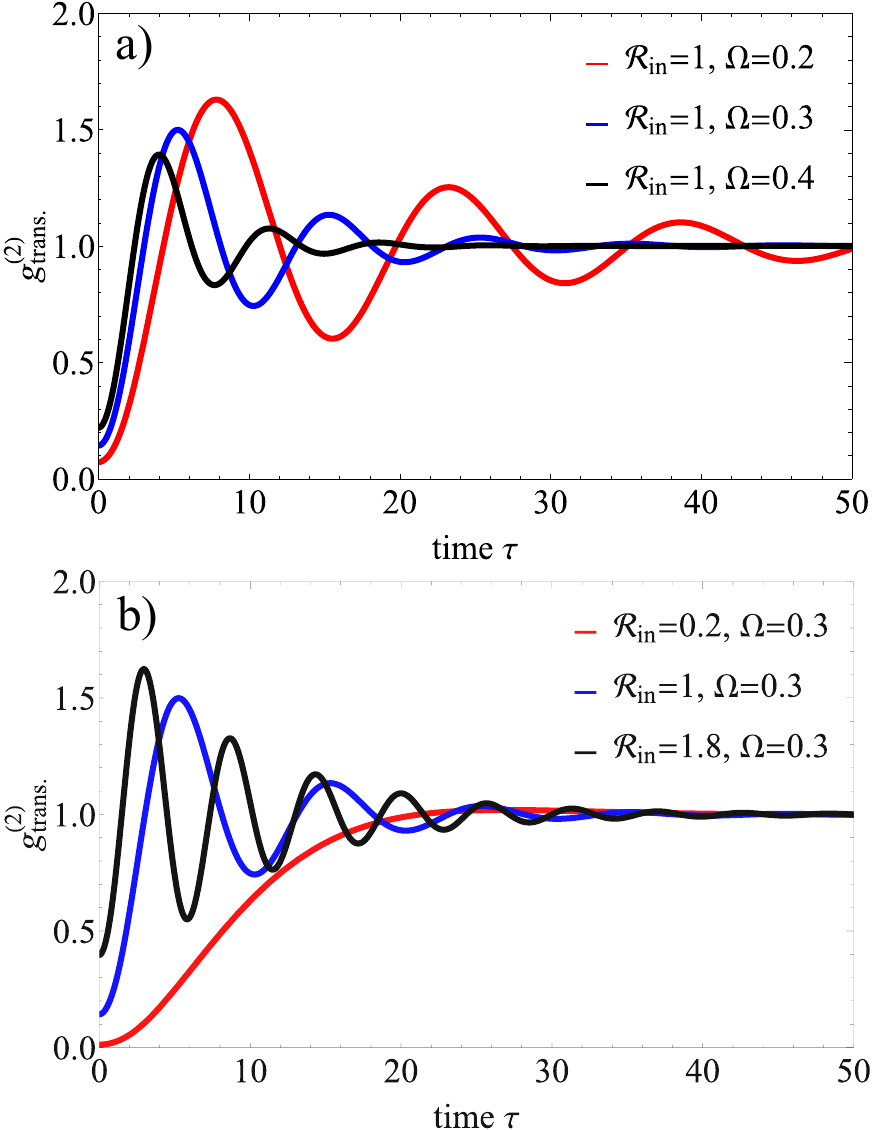}
\end{center}
\caption{Correlations in transmitted light under conditions of EIT. a) Numerical solution of Eqs.~(\ref{eq:3lv_1})-(\ref{eq:3lv_4}) shows slowing oscillations for descreasing $\Omega$. b) Fluctuations intensify both in frequency and in number with growing $\mathcal{R}_{\rm in}$.}
\label{fig:g2_full}
\end{figure}

As shown in Fig.~\ref{fig:g2_full}, the transmitted light is strongly antibunched and the photon correlation function $g^{(2)}_{\rm trans.}(\tau)$ exhibits damped oscillations at finite times. This reversed response as compared to the previously discussed case with a single exciton state, is readily understood by noting that the linear mirror is now completely transmissive, instead of being fully reflective. As a single photon is transmitted through the mirror it generates a Rydberg exciton for a time $t\sim\Delta\tau$, which blocks EIT for any other photons. More specifically, by preventing the formation of the EIT dark state, the Rydberg-exciton blockade exposes the strong photon coupling to the low-lying exciton states, which leads to high reflection and thereby reduces the simultaneous transmission of multiple photons.

Interestingly, we find that the degree of antibunching, $g^{(2)}_{\rm trans.}(0)$, depends on the amplitude $\mathcal{R}_{\rm in}$ of the driving field as well as the control Rabi frequency $\Omega$. We can analyze this behavior more systematically by first considering the limit of weak control fields, $\Omega\ll1$, in which we can adiabatically eliminate the dynamics of the intermediate states. Neglecting the time derivative in Eq.~(\ref{eq:P}) gives 
\begin{align}
 \hat{\mathcal{P}} = -\frac{2i}{\tilde{\Gamma}} \mathcal{R}_{\rm in} - \frac{2i\Omega}{\tilde{\Gamma}}\hat{\mathcal{S}},
\end{align}
and substitution into Eq.~(\ref{eq:S}) yields a single equation for the Rydberg exciton that can once again be mapped onto an effective spin-1/2 system. This makes it possible to obtain exact expressions for the steady-state exciton density 
\begin{align}\label{eq:elim_steady_state}
 \langle \hat{\mathcal{S}}^\dagger \hat{\mathcal{S}} \rangle &= \frac{\mathcal{R}_{\rm in}^2}{2 \mathcal{R}_{\rm in}^2+\Omega ^2}, 
\end{align}
and for the two-photon correlation of the transmitted light 
\begin{align}\label{eq:g2adiab}
 g^{(2)}_{\rm trans.}(\tau) = 1-\frac{e^{-\frac{3  \Omega ^2}{2} \tau}}{\sqrt{\kappa }} &\left[ 3 \Omega  \sinh \left(\frac{\sqrt{\kappa }  \Omega}{2}\tau  \right) \right. \\
 &\left. +\sqrt{\kappa } \cosh \left(\frac{\sqrt{\kappa } \Omega }{2}\tau \right) \right], \nonumber
\end{align}
where we have set $\Delta=0$ for simplicity and where $\kappa = \Omega ^2-16 \mathcal{R}_{\rm in}^2$. We see that the additional control-field coupling now makes it possible to independently tune the characteristic correlation time $\sim\Omega^{-2}$ and the oscillation frequency $\Omega\sqrt{-\kappa}$, by varying the control and probe field amplitudes, $\Omega$ and $\mathcal{R}_{\rm in}$. In particular, the weak-field limit $\Omega\ll1$ corresponds to strong photon correlations with persistent long-time oscillations, even at high probe-field intensities.

\begin{figure}[t]
\begin{center}
 \includegraphics[height=.3\textwidth]{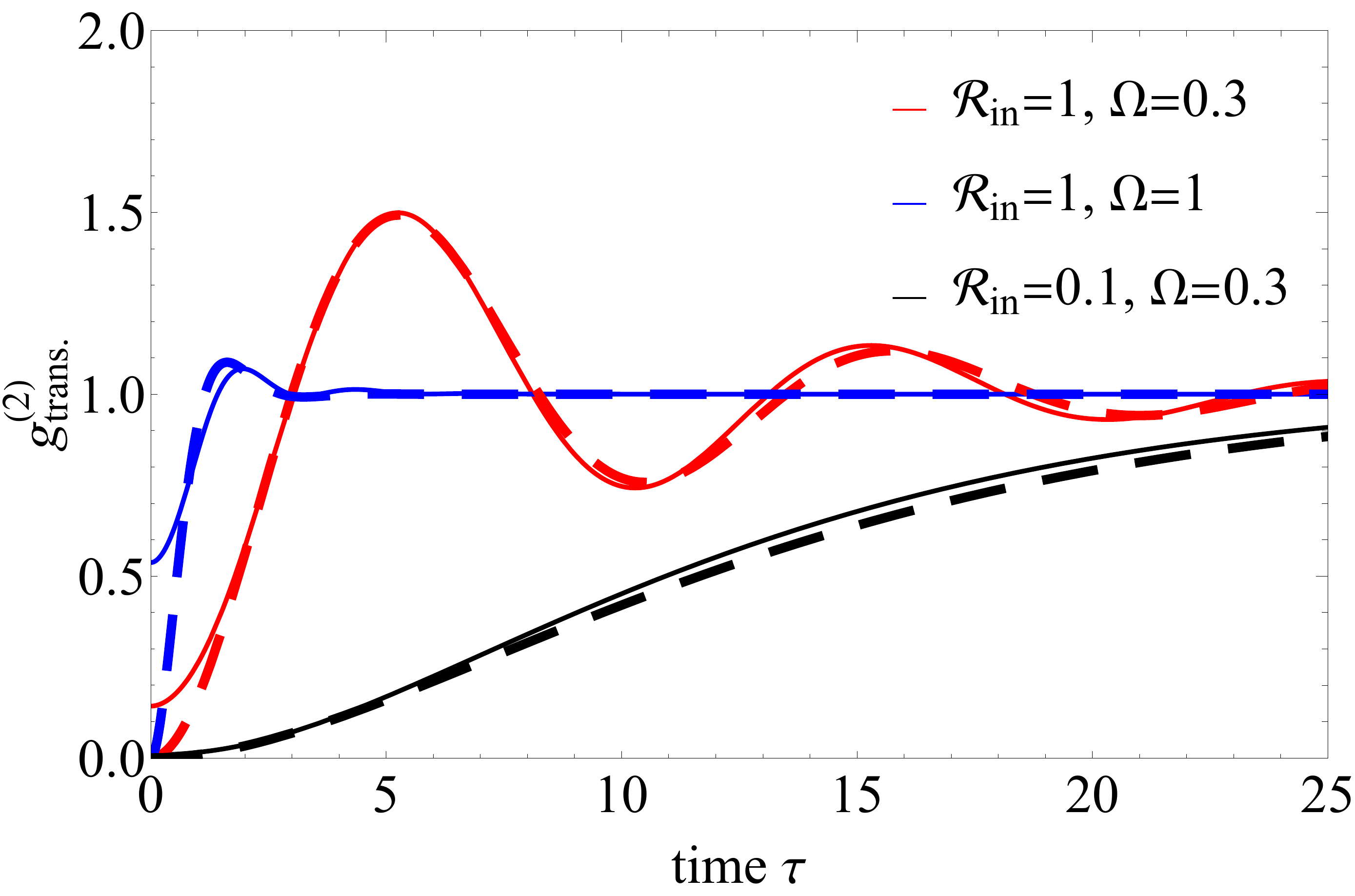}
\end{center}
\caption{Comparison of exact solution with adiabatic elimination. After large differences for small $\tau$, agreement is generally good.}
\label{fig:comp_elim}
\end{figure}

As shown in Fig.\ref{fig:comp_elim}, this adiabatic approximation yields a good description of emerging photon correlations for not too strong intensities of both applied laser fields, $\Omega,\mathcal{R}_{\rm in}<1$. We can gain a better understanding of the observed deviations at short times by considering the perturbative solution of Eqs.~(\ref{eq:3lv_1}-\ref{eq:3lv_4}) for small driving strengths, $R_{\rm in}\ll1$. The obtained Rydberg-exciton density 
\begin{align}
 \langle \hat{\mathcal{S}}^\dagger \hat{\mathcal{S}} \rangle = \frac{1}{\Omega^2} \mathcal{R}^2_{\rm in} +\frac{2 \left(\Omega^4-\Omega^2-1\right)}{\Omega^4 \left(\Omega^2+1\right)^2} \mathcal{R}_{\rm in}^4 + \mathcal{O} (\mathcal{R}^5_{\rm in})
\end{align}
establishes $\Omega\ll1$ as a condition for the adiabatic elimination and agrees with Eq.~(\ref{eq:elim_steady_state}) in this limit, while the perturbative expansion of the equal-time photon correlations 
\begin{align}
 &g_{\rm trans.}^{(2)}(0) = \ \frac{\Omega^4}{(1+\Omega^2)^2} \\
 &-\frac{8 \left(\Omega ^2 \left(11 \Omega ^6-39 \Omega ^4-108 \Omega ^2-36\right)\right)}{3 \left(\left(\Omega ^2+1\right)^3 \left(\Omega ^2+2\right)^2 \left(\Omega ^4+24 \Omega ^2+12\right)\right)} \mathcal{R}_\text{in}^2  \nonumber\\
 &\rightarrow \Omega^4+2 \mathcal{R}_\text{in}^2 ,
\end{align} 
shows that weak control and probe field amplitudes $\Omega$ and $\mathcal{R}_{\rm in}$ indeed permit generating strongly antibunched light as predicted in the adiabatic limit.

\begin{figure}[t]
\begin{center}
  \includegraphics[height=.5\textwidth]{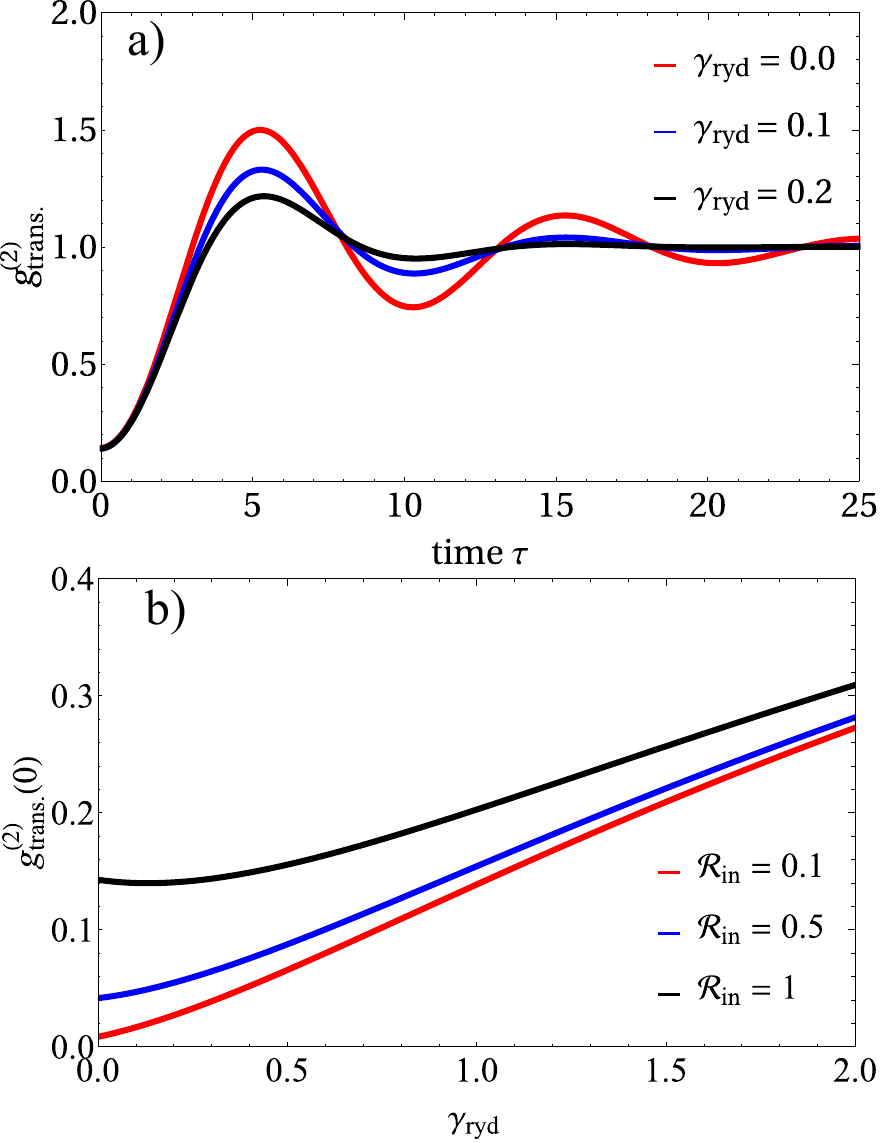}
\end{center}
\caption{Effects of Rydberg decay $\gamma_{\rm ryd}$ on light correlations. a) Examples of the correlation function for different $\gamma_{\rm ryd}$ at $\mathcal{R}_\text{in} = 1$ and $\Omega = 0.3$. b) Zero-time antibunching in the transmitted photons tends to decrease with $\gamma_{\rm ryd}$ and $\mathcal{R}_\text{in}$ but remains at remarkably high levels even in the presence of large Rydberg decay.}
\label{fig:decays}
\end{figure}

A final important factor is the linewidth of the excited exciton state. While the radiative Rydberg-state coupling is known to decrease with increasing principal quantum number, the influence of defects and non-radiative decay processes might remain substantial and limit the linewidth $\gamma_{\rm ryd}$ of the excited state. We can investigate such dissipation effects on the resulting photon correlations by including a decay term, $-\tfrac{\gamma_{\rm ryd}}{2}\hat{\mathcal{S}}$, in Eq.~(\ref{eq:S}). In Fig.\ref{fig:decays}a we show the obtained two-photon correlation of the transmitted light for different values of $\gamma_{\rm ryd}$. Surprisingly, the short-time behavior of the photon correlations remains virtually unaffected by excited-state decoherence, even for values of $\gamma_{\rm ryd}=0.2\gamma$ that are already twice as large as the non-radiative linewidths that have been measured for ground-state excitons in TMDC monolayers \cite{back2018,scuri2018}. As shown if Fig.\ref{fig:decays}b, it turns out that the excited-state decay rate can be substantially larger than this value and still retain significant antibunching of the transmitted light. This surprising level of robustness of the generated photon correlations against Rydberg-state broadening can be understood intuitively from the fact that -- even in the presence of additional broadening -- the control-field coupling to the excited state will always lower the otherwise near-perfect reflectivity generated on the lower exciton transition, driven by the probe field $R_{\rm in}$. Hence, the interaction blockade of the excited-state excitons can still provide an efficient nonlinear switching mechanism of the monolayer transmission and generate strong photon correlations despite substantial excited-state broadening that may exceed the decay rate of the low-lying exciton state. Provided that the interaction blockade remains effective, the asymptotic solution of the equal-time correlation function ($\gamma_\text{ryd} \gg 1, \Omega$ and $\mathcal{R}_\text{in}\ll 1$)
\begin{align}
 g_{\rm trans.}^{(2)}(0) \approx 1-\frac{4}{\gamma_\text{ryd}} + \mathcal{O} (\gamma_\text{ryd}^{-2}),
\end{align}
suggests that much larger decay rates on the order of $\gamma_{\rm ryd}\sim10\gamma$ still permit the generation of correlated, nonclassical light with photon antibunching well below current values of $g^{(2)}(0)\sim0.95$ in semiconductor microcavities \cite{bloch2018,delteil2019}.

\section{Conclusion}
In this work, we have elucidated the effects of finite-range exciton interactions on the optical properties of atomistically thin mirrors formed by two-dimensional semiconductors. Remarkably, this combination turned out to permit an exact solution of the underlying many-body problem of interacting excitons coupled to quantum light fields. This stands in marked contrast to equivalent bulk systems \cite{walther2020} or cavity settings \cite{jia2018}, where a theoretical description \cite{sevincli2011,gorshkov2011,gorshkov2013,Zeuthen2017,bienas2020,georgakopoulos2018} beyond the few-photon or semiclassical limits remains a formidable numerical challenge. We have made use of this property to investigate the properties of the scattered light and showed that the interaction-induced nonlinear reflection and transmission of the semiconductor can generate highly nonclassical states of light. We have proposed a two-photon coupling scheme that permits exploiting the strong photon-coupling to ground-state excitons, while a classical control beam is used to efficiently couple these excitons to an interacting excited state. By realizing conditions of electromagnetically induced transparency, this approach provides an efficient nonlinear switching mechanism between high transmission and high reflection and, thereby, to convert a coherent input field into strongly antibunched photons. Importantly, the proposed two-photon scheme is surprisingly robust against unavoidable line broadening of the excited Rydberg state, which offers a promising outlook on future experiments. 

While experiments on pristine samples of Cu$_2$O \cite{kazimierczuk2014giant}, have already revealed high-lying Rydberg states with strong interactions \cite{walther2018interactions} and sizable nonlinear signals \cite{walther2020}, equivalent observations for two-dimensional excitons are currently limited to lower lying states. Measurements on monolayer TMDCs have observed excited states of excitons  \cite{chernikov2014exciton,stier2018,gu2019,wang2020ryd} with assigned principal quantum numbers of up to $n=11$ and found signatures for the enhancement of the induced optical nonlinearities and exciton interaction range with increasing principal quantum number \cite{gu2019}. A recently measured blockade radius of $\sim25$~nm at $n=2$ in WSe$_2$ monolayers \cite{gu2019} suggests blockade radii of $\sim1$~$\mu$m for moderate principal quantum numbers of $n\sim5...10$. Small beam waists of $\sim1$~$\mu$m are possible with optical fibers \cite{brambilla2010} and the present calculations demonstrate that significant antibunching below previous measurements \cite{bloch2018,delteil2019} should still be possible for blockade radii that are 4-6 times smaller than this value, while fabricated masks and electrostatic gate control \cite{mak2013, wang2018} may be used to isolate small excitation spots that enable complete interaction blockade for even smaller principal quantum numbers. 

Such coupling of focused in- and outgoing light via proximate fibre ends also permits creating high-quality optical resonators \cite{besga2015} that can lead to transverse confinement of optical modes, which was recently shown to generate observable photon correlations with GaAs quantum wells \cite{bloch2018,delteil2019}. The combination of strong mode confinement, the mode-selective photon coupling of TMDC monolayers in an optical resonator \cite{zeytinoglu2018,Wild2018,zhou2020} and the nonlinear mechanisms described in this article, thus presents a promising approach to quantum nonlinear optics in the solid-state that remains to be explored in future work. Here, the remarkable electro-optical properties of TMDC monolayers open up a number of interesting  possibilities. Their strong spin-orbit coupling, for example, leads to valley-dependent photon coupling with tunable polarization selection \cite{xiao2012coupled}. This, in turn, can generate a polarization-dependent nonlinearity that may give rise to polarization entanglement and offers an enabling mechanism for generating and manipulating more complex quantum states of light.

\section{Acknowledgments}
We thank Nikola \v{S}ibali\'c, Nikolaj Sommer J\o{}rgensen and Trond Andersen for useful discussions. This work has been supported by the EU through the H2020-FETOPEN Grant No. 800942640378 (ErBeStA), by the DFG through the SPP1929, by the Carlsberg Foundation through the Semper Ardens Research Project QCooL, by the DNRF through the Center for Complex Quantum Systems (Grant agreement no.: DNRF156), and by the NSF through a grant for the Institute for Theoretical Atomic, Molecular, and Optical Physics at Harvard University and the Smithsonian Astrophysical Observatory. SFY would like to thank for support from the NSF through the CUA PFC (context of Rydberg 2D arrays) and from the DOE through group grant DE-SC0020115 (for applications to excitons). 

\section*{References}

% \bibliographystyle{apsrev4-1}
% \bibliography{NonLinXMirror}

%

\end{document}